\documentclass[11pt,dvips]{article}

\usepackage{epsfig,times} 
\usepackage{picinpar}
\usepackage{wrapfig}
\usepackage{floatflt}
\makeatletter
\renewcommand{\section}{\@startsection{section}{1}{0in}
	{0.4\baselineskip}{0.1\baselineskip}{\Large\bf}}
\renewcommand{\subsection}{\@startsection{subsection}{2}{0in}
	{0.25\baselineskip}{-\baselineskip}{\large\bf}}
\renewcommand{\subsubsection}{\@startsection{subsubsection}{3}{0in}
	{0.1\baselineskip}{-\baselineskip}{\normalsize\bf}}
\makeatother
\begin{document}

%
\makeatletter\newcommand{\ps@icrc}{
\renewcommand{\@oddhead}{\slshape{OG.3.1.10}\hfil}}
\makeatother\thispagestyle{icrc}
%
%

\begin{center}
%
{\LARGE \bf Superluminal Particles 
in Cosmic-Ray Physics}
\end{center}

\begin{center}
%
%
{\bf L. Gonzalez-Mestres$^{1,2}$}\\
{\it $^{1}$Laboratoire de Physique Corpusculaire, Coll\`ege de France, 75231 Paris Cedex 05, France\\
$^{2}$L.A.P.P., B.P. 110, 74941 Annecy-le-Vieux Cedex, France}
\end{center}

\begin{center}
{\large \bf Abstract\\}
\end{center}
\vspace{-0.5ex}
%
%
Present low-energy bounds on Lorentz symmetry violation do not
allow to exclude the possible existence of superluminal particles
({\it superbradyons}) with critical speed in vacuum 
$c_i~\gg ~c$ ($c$ = speed of
light) whose kinematical properties would be close to those of "ordinary"
particles (bradyons) 
apart from the difference in critical speed. If they exist,
superbradyons may be the basic building blocks of vacuum and matter at
Planck scale, provide most of the matter in the Universe and be natural
dark matter candidates. We present an updated discussion of their
theoretical and experimental properties, especially as cosmic ray
primaries or sources, as well as problems related to their possible
direct detection.
%

\vspace{1ex}

%
%
\section{The Meaning of Relativity}
\label{relativity.sec}

Lorentz symmetry, viewed as a property of dynamics, 
implies no reference to absolute
properties of space and time 
(Gonzalez-Mestres, 1995a). In a two-dimensional
galilean space-time,
the wave equation:
\equation
\alpha ~\partial ^2\phi /\partial t^2~-~\partial ^2\phi /\partial x^2 = F(\phi )
\endequation
\noindent
with $\alpha$ = $1/c_o^2$ and $c_o$ = critical
speed, remains unchanged under "Lorentz" transformations leaving
invariant the squared
interval
$ds^2 = dx^2 - c_o^2 dt^2$ . 
Matter made with solutions of equation (1)
would feel a relativistic space-time even if the real space-time is 
galilean and if an absolute rest frame exists in the
underlying dynamics beyond the wave equation.
The solitons of the sine-Gordon equation are
obtained taking in (1):
\equation
F(\phi )~~ = ~~-~(\omega /c_o)^2~sin~\phi
\endequation
\noindent
$\omega $ being a characteristic frequency of the dynamical system.
The two-dimensional universe made of such sine-Gordon solitons 
would behave like a two-dimensional minkowskian
world with the laws of special relativity.
The actual structure of space and time
can only be found by going 
to deeper levels of resolution where the equation fails,
similar to the way
high-energy accelerator experiments explore the inner structure of
"elementary" particles (but cosmic rays have the highest attainable
energies).
In such a scenario, that cannot be ruled out by any present experiment,
superluminal sectors of matter can exist and even be its ultimate
building blocks. This clearly makes sense, as:
a) in a
perfectly transparent crystal, at least two critical speeds can be
identified, those of light and
sound; b) the potential approach to 
lattice dynamics in solid-sate physics
is precisely the form of electromagnetism
in the limit $c_s~c^{-1}~\rightarrow ~0$ , where $c_s$ is the speed of sound
and $c$ that of light.
Superluminal sectors of matter can be consistently
generated (Gonzalez-Mestres, 1995a and 1996) replacing in the Klein-Gordon 
equation the
speed of light by a new critical speed $c_i$ $\gg $ $c$
(the subscript $i$ stands for the $i$-th superluminal sector). All
standard kinematical concepts and
formulas remain correct, leading to particles with
positive mass and energy ({\bf superbradyons}) which are not tachyons.
The energy $E$ and momentum $p$
of a superbradyon with mass $m$ and critical speed $c_i$
will be given by the generalized relativistic equations:
\begin{eqnarray}
p~=~m~v~(1~-~v^2c_i^{-2})^{-1/2}~ \\
~E~=~m~c_i^2~(1~-~v^2c_i^{-2})^{-1/2} \\
E_{rest}~~=~~m~c_i^2~~~~~~~~~~~~~~~
\end{eqnarray}
\noindent
where $v$ is the speed and $E_{rest}$ the rest energy.
Each superluminal sector will have its own Lorentz invariance
with $c_i$ defining the metric, and is expected to generate a sectorial
"gravity". We call the sector made of particles with critical speed in
vacuum = $c$ the "ordinary" sector (made of "ordinary" particles
or "bradyons"). Being able to write  
formulae (3)-(5) for all sectors simultaneoulsy requires the
existence of an "absolute" rest
frame, ({\bf the vacuum rest frame}, VRF, perhaps close to that
suggested by the study of cosmic microwave background radiation), 
the only one where this will be 
possible.
Furthermore, interactions between two different
sectors will break both Lorentz invariances and deform their 
kinematics (an example is given in Gonzalez-Mestres, 1997a; 
for more recent material,
see Gonzalez-Mestres, 1998, and references therein).   

The dynamical links between different sectors, including between the 
"ordinary" and the superluminal sectors, are expected to occur at
very high energy and very short distances, in relation with new physics
generated above the relevant fundamental length(s) (Gonzalez-Mestres,
1995b and 1997b). Such an interaction is also relevant to the very early
Universe and is expected to considerably modify its scenario. 
The critical energy scales will naturally be associated to
critical temperatures, leading to a "multi-transition" early Universe where
matter itself will transform (Gonzalez-Mestres,
1995b and 1997b). But, at low energy, none of these effects will be 
apparent in standard tests of special relativity.
At the fundamental lenght scale, and taking a simplified
two-dimensional illustration, gravitation may even be a composite
phenomenon (Gonzalez-Mestres, 1997b), related for instance to
fluctuations of the parameters of equations like:
\equation
A~d^2/dt^2~~[\phi ~(n)]~+~H~d/dt~[~\phi ~(n+1)~-~\phi ~(n-1)]~-~\Phi ~[\phi ]~=~0
\endequation
where we have quantified space to schematically account
for the existence of the fundamental length $a$, $\phi$ is a wave function, 
$n$ designs by an integer lattice sites spaced by a distance 
$a$, $A$ and $H$ are
coeficients and
$\Phi ~[\phi ]$ is defined by:
\equation
\Phi ~[\phi ]~=~K_{fl}~[2~\phi ~(n)~-~\phi 
~(n-1)~-~\phi ~(n+1)]~+~\omega _{rest}^2~\phi 
\endequation
$K_{fl}$ being a coefficient and $(2\pi )^{-1}~\omega _{rest}$ a rest frequency. 
In the continuum limit, the coefficients $A~=~g_{00}$ , $H~=~g_{01}~
=~g_{10}$ and $-K_{fl}~=~g_{11}$ can be regarded as the matrix elements of
a space-time bilinear metric with equilibrium values: $A~=~1$ , $H~=~0$ and
$K_{fl}~=~K$~. Then, a small local fluctuation:
\equation
A~=~1~+~\gamma
\endequation
\equation
K_{fl}~=~K~(1~-~\gamma )
\endequation
with $\gamma ~\ll ~1$ would be equivalent to a small, static
gravitational field
created by a far away source. With our deformed Lorentz symmetry 
approach to relativity (see paper HE.1.3.16 of these
Proceedings), general relativity would naturally be preserved at
low energy.

\section{Superluminal Matter}
\label{superl.sec}
Apart from the difference in critical speed, superbradyons can be
rather "normal" objects to which generalized field theories can possibly
be applied
(Gonzalez-Mestres, 1997a and 1998), the definition of causality 
and cosmological horizon depending
on the critical speeds under consideration.

\subsection{Superbradyons in the Universe}
\label{superb.sec}
If superbradyons are the
building blocks of "ordinary" matter, and $c$ is just a composite
critical speed like the speed of sound, we expect to find new (superluminal)
forms of matter beyond the Planck scale (where "ordinary" particles
may cease to exist), as well as a new physics up to
much higher energies. At the same time, superbradyons may well be able to
propagate in our vacuum at lower energies, just like light can propagate 
in a transparent crystal. We expect them to emit "Cherenkov radiation"
(ordinary particles and other superbradyons with a lower critical speed
in vacuum) when
they propagate in vacuum at very high speed. 
Primaries, secondaries and decay products of
this radiation may in principle have extremely high energies and reach 
suitable detectors. Although superbradyons should eventually loss their 
kinetic energy
until they get traveling at a speed lower than light, if
superluminal matter is very abundant it can be continuously generating
new fast superbradyons from the decay of very heavy ones. 
Superbradyons may generate alternatives
to inflation in a "Big Bang"-like cosmology (Gonzalez-Mestres, 1995b)
and provide nowadays most of the matter in the Universe in extended
Friedmann models (Gonzalez-Mestres, 1997b)
incorporating several sectors of matter.
As a rough example, 
we assume that a theory of all gravitation-like forces can be
built, taking at each point the
vacuum rest frame, and generalize Friedmann equations writing for
a flat Universe in the present epoch (where
pressure can be neglected):
\equation
R^{-1}~d^2R/dt^2~~\approx ~~- 4\pi ~ Z_2~Z_1^{-1}/3~+~\Lambda /3
\endequation
\equation
(R^{-1}~dR/dt)^2~~\approx ~~ 8\pi ~Z_2~Z_1^{-1}/3~+~\Lambda /3
\endequation
where $R$ is the distance scale,
$\Lambda $ the cosmological constant and, in a simplified
scheme:
\equation
Z_1~~=~~\rho _a~+~\rho _O ~+~\Sigma _i~(\rho _{a,i}~+~
\rho _{O,i})~
\endequation
\equation
Z_2~~=~~G_a~\rho _a^2~+~G_O~\rho _O ^2~+~\Sigma _i~(G_{a,i}~\rho _{a,i}^2~+~
G_{O,i}~\rho _{O,i}^2)
\endequation
\noindent
where $G_a~=~G_N$ is Newton's gravitational constant,
$\rho _a$ the density of "acoustic" ordinary matter (the "acoustic" band of 
the bradyon spectrum as composite objects at Planck scale, i.e. the
conventional "elementary" particles),
$\rho _O$ the density of "optical" ordinary matter 
(taken to be positive, see Gonzalez-Mestres, 1997b), $\rho _{a,i}$ and
$\rho _{O,i}$ the densities of "acoustic" and "optical" matter of the $i$-th
superluminal sectos (again, taking the densities of "optical" particles
to be positive, see Gonzalez-Mestres, 1997b), 
and the $G$'s are effective gravitation-like
coupling constants.
$Z_2~Z_1^{-1}$ replaces the usual expression $G_N~\rho $ in standard
Friedmann equations.
$Z_1$ is the total density of "particle matter",
where the expression "particle matter"
designs all
possible excitations of vacuum that we can describe as particles.
These expressions can be derived, for instance, by associating
standard Friedmann equations (with only one "gravitational" component)
to a lagrangian in terms of $R$ and $dR/dt$ , and generalizing the
expressions for kinetic and potential energies in the limit where gravitational
couplings between different components of $Z_1$ are small. An interaction
between the different "gravitational" components of $Z_1$ is,
even in this case,
implicitly generated by the constraint that $R$ and $dR/dt$ are space-time
variables common to all the kinds of matter we consider.
Since, at the same time, cosmology considers space-time as being
generated by matter, this is indeed an effective dynamical interaction
between matter from different sectors. The role of
vacuum, the ground state of cosmic matter, is crucial in the generation of a 
single, absolute space-time
with a local absolute rest frame.

Accelerator experiments at future machines (LHC, VLHC...) can be a way
to search for superluminal particles
(Gonzalez-Mestres, 1997a, 1997c and 1998). 
However, this approach is limited by
the attainable energies, luminosities, signatures and background levels.
The coupling of ordinary particles to the superluminal sectors is
expected to be strongly energy-dependent.  
Although direct searches at accelerators provide unique chances and
must be carried on, they 
will only cover a small domain of the allowed parameters
for superluminal sectors of matter. Cosmic-ray experiments
are not limited in energy and naturally provide very low background levels:
they therefore allow for a more general and, on dynamical grounds,
better adapted exploration of both Lorentz symmetry violation and
possible superluminal particles. 

It must also be realized that, if the Poincar\'e
relativity principle is violated, a $1~TeV$ particle cannot be turned into
a $10^{20}~eV$ particle of the same kind by a Lorentz transformation, and
collider events cannot be made equivalent to cosmic-ray events. Very-forward
experiments at very high-energy accelerators, covering the kinematical
domain equivalent (according to special relativity) to very high-energy
cosmic-ray events, would, if technically feasible, be an extremely
important research domain allowing to directly test Lorentz symmetry in
unexplored kinematical regions. This line of research would in
principle justify building accelerators (e.g. for $p - p$ collisions)
at energies as high as $\approx
~400~TeV$ per beam 
(equivalent to $\approx ~3.10^{20}~eV$ cosmic proton hiting a proton
at rest in the detector). Such an accelerator energy
would simultaneously open a new
window to direct search for superbradyons.
It must be noticed that the highest observed cosmic-ray energies 
are closer to Planck scale ($\approx ~10^{28}~eV$) than to
electroweak
scale ($\approx ~10^{11}~eV$): therefore, if Lorentz symmetry is violated,
the study of the highest-energy cosmic rays
provides a unique microscope directly focused on Planck scale.
The search for very rare events due to
superluminal particles in AUGER, AMANDA, OWL, AIRWATCH FROM SPACE...
can be a crucial ingredient of this unprecedented
investigation (Gonzalez-Mestres, 1997d), searching for the ultimate
components of matter beyond Lorentz symmetry. 

\subsection{Experimental Considerations}
\label{experim.sec}

In what follows, we neglect Lorentz symmetry violation for the
physical processes governing our experimental set-up and assume
that the earth is not moving at relativistic speed with respect to the local
vacuum rest frame. With these hypothesis, superbradyon kinematics has
been discussed at length in (Gonzalez-Mestres, 1997a and 1998).
Attention must be paid, for instance, to timing signatures: a 
superbradyon moving with velocity ${\vec {\mathbf v}}_{\mathbf i}$
with
respect to the VRF,
and emitted by an astrophysical object, can reach an observer
moving with laboratory
speed  ${\vec {\mathbf V}}$ in the VRF at a time, as
measured by the observer, previous to the emission time.
This remarkable astronomical phenomenon will
happen if ${\vec {\mathbf v}}_{\mathbf i}.{\vec {\mathbf V}}~>~c^2$~, and the
emitted particle will be seen to evolve backward in time (but it evolves
forward in time in the VRF, so that again the reversal of the time
arrow is not really a physical phenomenon). 
Our study also points out that, for $V~\ll ~c$ and
${\vec {\mathbf v}}_{\mathbf i}. {\vec {\mathbf V}}~\gg ~c^2$ , the speed
${\vec {\mathbf v}}{\mathbf {_i}'}$ of the superbradyon as seen by the 
observer tends to the limit
${\vec {\mathbf v}}{\mathbf {_i}{^\infty }}$ , where:
\equation
{\vec {\mathbf v}}{\mathbf {_i}{^\infty }}~
({\vec {\mathbf v}}_{\mathbf i})~~=~~-~{\vec {\mathbf v}}_{\mathbf i}~c^2~
({\vec {\mathbf v}}_{\mathbf i}.{\vec {\mathbf V}})^{-1}
\endequation
which sets a universal high-energy limit, independent of $c_i$ ,
to the speed of superluminal particles as
measured by ordinary matter in an inertial rest frame other than the
VRF. This limit is not isotropic, and depends on the angle between
the speeds ${\vec {\mathbf v}}_{\mathbf i}$ and ${\vec {\mathbf V}}$ .
A typical order of magnitude for
${\vec {\mathbf v}}{\mathbf {_i}{^\infty }}$ on earth is
${\vec {\mathbf v}}{\mathbf {_i}{^\infty }}~\approx 10^3~c$ if
the VRF is close to
that suggested by cosmic background radiation.
Furthermore, for a very high-speed
superluminal cosmic ray with critical speed $c_i~\gg ~c$ ,
it turns out (Gonzalez-Mestres, 1997a and 1998)
that the momentum, as measured in the laboratory,
does not provide directional
information on the source, but on the VRF.
Velocity provides directional information on the source,
but can be measured only if the
particle interacts several times with the detector, 
or if 
photons or neutrinos are emitted simultaneously.

Annihilation of pairs of cosmic superluminal particles into ordinary or
superluminal ones can
release very large kinetic energies.
Decays of superbradyons may play a similar role, as well as 
collisions (especially, inelastic with very large energy
transfer) of high-energy superluminal particles with
extra-terrestrial ordinary matter.
Superluminal particles moving at $v_i~>~c$ can release anywhere the
above mentioned "Cherenkov"
radiation in vacuum,
providing a new source of (superluminal or ordinary)
high-energy cosmic rays.
High-energy superbradyons
can directly reach the earth. Their interactions have been discussed in
(Gonzalez-Mestres, 1996, 1997a, 1997d and 1998).   
Signatures of very high-energy superluminal particles seem strong enough to 
escape all backgrounds. Superluminal sources escape quite generally
(Gonzalez-Mestres, 1996 and 1998) the 
Greisen-Zatsepin-Kuzmin cutoff (Greisen, 1966; Zatsepin and Kuzmin, 1966).  

%
%
%
\vspace{1ex}
\begin{center}
{\Large\bf References}
\end{center}
%
Gonzalez-Mestres, L., 1995a, Proceedings of the
Moriond Workshop on "Dark Matter in Cosmology, Clocks and Tests of
Fundamental Laws", Villars January 1995 ,
Ed. Fronti\`eres, p. 645.\\
Gonzalez-Mestres, L., 1995b,
Proceedings of the IV International Workshop on Theoretical and
Phenomenological Aspects of Underground Physics
(TAUP95), Toledo September 1995, Ed. Nuclear Physics Proceedings, p. 131.\\
Gonzalez-Mestres, L., 1996, contribution to the 28$^{th}$
International Conference on High Energy Physics (ICHEP 96), Warsaw July 1996,
paper hep-ph/9610474 of LANL (Los Alamos) electronic archive.\\
Gonzalez-Mestres, L., 1997a, papers physics/9702026
and physics/9703020 of LANL archive.\\
Gonzalez-Mestres, L., 1997b, paper
physics/9704017 of LANL archive.\\
Gonzalez-Mestres, L., 1997c, contribution to the Europhysics
International Conference on High-Energy Physics
(HEP 97), Jerusalem August 1997, paper physics/9708028.\\
Gonzalez-Mestres, L., 1997d, Proc. 25th ICRC (Durban, 1997), Vol. 6, p. 109.\\
Gonzalez-Mestres, L., 1998, Proc. "Workshop on Observing
Giant Cosmic Ray Air Showers From $>$ 10$^{20}$ Particles From Space", College
Park, November 1997, AIP
Conference Proceedings 433, p. 418.\\
Greisen, K., 1966, Phys. Rev. Lett. 16, 748.\\
Zatsepin, G.T., \& Kuzmin, V.A., 1966, Pisma Zh. Eksp. Teor. Fiz. 4, 114. 
\end{document}